\documentclass[11pt]{article}

\usepackage{amsmath}
\usepackage{amssymb}
\usepackage{latexsym}
\usepackage{amsfonts}
\usepackage{amsthm}
\usepackage{bbm,dsfont}

\usepackage{color}
\usepackage{graphicx}
\usepackage{graphics}
\usepackage[all,cmtip]{xy}

\newtheorem{theorem}{Theorem}

\theoremstyle{definition}
\newtheorem{example}{Example}

\newcommand{\C}{\mathbb C} 
\newcommand{\R}{\mathbb R} 
\newcommand{\N}{\mathbb N} 

\newcommand{\hi}{\mathcal H} 

\newcommand{\sh}{\mathcal{S(H)}} 
\newcommand{\eh}{\mathcal{E(H)}} 
\newcommand{\ip}[2]{\langle {#1}|{#2}\rangle} 
\newcommand{\tr}[1]{\mathrm{tr}\left[ {#1} \right]} 
\newcommand{\kb}[2]{|#1\,\rangle\langle\,#2|} 
\newcommand{\pee}[1]{P[#1]} 
\newcommand{\fii}{\varphi}
\newcommand{\veps}{\varepsilon}
\newcommand{\eps}{\epsilon}

\newcommand{\stfm}{{\mathcal I}} 
\newcommand{\sfq}{\mathsf{Q}}
\newcommand{\sfp}{\mathsf{P}}
\newcommand{\qhat}{Q}
\newcommand{\phat}{P}
\newcommand{\id}{\mathbbm{1}} 
\newcommand{\nul}{\mathbbm{O}} 

\newcommand{\br}{\mathcal{B}(\R)} 
\newcommand{\brr}{\mathcal{B}(\R^2)} 
\newcommand{\prob}{\mathsf{p}} 

\newcommand{\E}{\mathcal{E}}

\newcommand{\app}{\mathcal{A}}

\newcommand{\W}{{\mathcal W}}

\newcommand{\delq}[1]{\W_{\varepsilon_1,\delta}({#1},\sfq)}
\newcommand{\delp}[1]{\W_{\varepsilon_2,\delta}({#1},\sfp)}
\newcommand{\deq}[1]{\W_{\varepsilon_1}({#1},\sfq)}
\newcommand{\dep}[1]{\W_{\varepsilon_2}({#1},\sfp)}

\newcommand{\iqd}{J_{q;\delta}}

\newcommand{\iqw}{J_{q;w}}

\begin{document}

\title{\large\sc ``No Information Without Disturbance":\\
Quantum Limitations of Measurement}

\author{Paul Busch\thanks{Electronic mail: pb516@york.ac.uk}\\
{\small Perimeter Institute for Theoretical Physics, Waterloo, ON, Canada}\\
{\small Permanent address: Department of Mathematics, University of York, York,  UK}
}
\date{\small June 2007}

\maketitle

\begin{abstract}
\noindent In this contribution I review rigorous formulations of
a variety of limitations of measurability in quantum
mechanics. To this end I begin with a brief presentation of the conceptual
tools of modern measurement theory. I will make precise the notion
that quantum measurements necessarily alter the system under
investigation and elucidate its connection with the complementarity
and uncertainty principles.
\end{abstract}

\section{Introduction}

It is a great honor and pleasure for me to contribute to this
celebration of the scientific life work and achievements of Abner
Shimony, from whom I have received much inspiration, personal
encouragement and the gift of friendship in a decisive period of my
scientific career. When I came to know Abner more closely, I was
thrilled to realize the close  agreement between our quantum
mechanical world views; and ever since, when contemplating foundational issues,
I found myself often wonder: ``What would Abner say?". I am proud to share with 
Abner one piece of work on an important item of ``unfinished business", a paper 
on the insolubility of the quantum
measurement problem \cite{BuSh96}, which I hope may prove useful as a 
stepping stone towards resolving this problem. In this
contribution I will address another area of concern to Abner, one
that remains even when the measurement problem is suspended: quantum
limitations of measurements.

By way of introduction of terminology and notation I briefly review the basic
and most general probabilistic structures of quantum mechanics, encoded
in the concepts of states, effects and observables; I then recall how these
objects enter the modeling of measurements (Section \ref{sec:mmt}).

This general framework of quantum measurement theory will then be used
to obtain precise formulations and proofs of some long-disputed
limitations of quantum measurements, such as the inevitability of disturbance
and entanglement in a measurement, the impossibility of repeatable measurements 
for continuous quantities, and the  incompatibility between conservation
laws and the notion of repeatable sharp measurements (Section \ref{sec:qu-lim}).
In Section \ref{sec:CU}
the {\em ``classic"} quantum limitations expressed by the complementarity and
uncertainty principles are revisited. Appropriate operational measures of
inaccuracy and disturbance for the formulation of quantitative trade-off relations for
(joint) measurement inaccuracies and disturbances have been introduced
in recent years; these will be discussed in  Section \ref{sec:inacc-dist}.

I conclude  with an outlook on open questions (Section \ref{sec:conclusion}).

\section{Quantum Measurement Theory - Basic Concepts}\label{sec:mmt}

\subsection{States, effects and observables}

Every
quantum system is represented by a finite or infinite-dimensional,
separable Hilbert space $\hi$ over the complex field $\C$. States
are described as positive operators\footnote{The term operator will
be taken as shorthand for ``linear operator". With $A\le B$ or
equivalently $B\ge A$ we denote the usual ordering of self-adjoint
operators; thus, $A\le B$ if and only if
$\ip\fii{A\fii}\le\ip\fii{B\fii}$ for all $\fii\in\hi$. An operator
$A$ is {\em positive} if $A\ge \nul$, the null operator.} $T$ of
trace equal to one.\footnote{We remark that our notation follows
closely that of the monograph \cite{QTM}. The letter $T$ was chosen
there to denote a state since it is the first letter of the Finnish
word for ``state"; the authors of that monograph found this
preferable to $W$, which would stand for the German word for
``knowledge", or $\rho$, which is reminiscent of the phase space
density with its classical connotations. Linguistic balance between
the authors was maintained by taking $Z$ to denote the pointer
(``Zeiger") observable in a measurement scheme (see below).
Naturally, $\mathcal M$ will stand for the  English term
``measurement".} The set of states $\sh$ is a convex subset of the
real vector space of all self-adjoint trace-class operators. The
role of a quantum state is to assign a probability to the outcome of
any measurement; in other words, associated with every measurement
with possible outcomes $\omega_i$, $i=1,2,\dots$, are mappings
$\E_i:\sh\to[0,1]$ assigning the probabilities
$\prob_T(\omega_i)\equiv \E_i(T)$. Since mixtures of states lead to
the corresponding mixtures of probabilities, it follows that the
mappings $\E_i$ are affine and hence extend uniquely to bounded
positive linear mappings. Since the dual space of the trace class is
isomorphic to the vector space of bounded operators, each $\E_i$ is
of the form $\E_i(T)=\tr{TE_i}$, where $E_i$ is an operator
satisfying $\nul\le E_i\le\id$ (here $\id$ denotes the identity
operator). Such operators are called {\em effects}. The set of
effects will be denoted $\eh$. The normalization of the probability
distributions $\prob_T$ ($\sum_i\prob_T(\omega_i)=1$) entails the
condition
\begin{equation}\label{eqn:normaliz}
\sum_i E_i=\id.
\end{equation}

The mapping $\omega_i\mapsto E_i$ together with the property
(\ref{eqn:normaliz}) is a (discrete) instance of a normalized
positive-operator-valued measure (POVM), the general definition
being that of an operator-valued mapping $X\mapsto E(X)$ with the
following properties: (i) the domain consists of all elements $X$ of
a $\sigma$-algebra $\Sigma$ of subsets of an outcome space
$\Omega$;
(ii) the operators $E(X)$ in the range are effects; (iii) the
mapping is $\sigma$-additive (with infinite sums defined as weak
limits): $E(\bigcup_iX_i)=\sum_iE(X_i)$ for any finite or countable
family of mutually disjoint sets in $\Sigma$; (iv) $E(\Omega)=\id$.
POVMs are taken as the most general representation of an {\em
observable}. In this contribution the measurable space of outcomes
$(\Omega,\Sigma)$ will be $(\R,\br)$ or $(\R^2,\brr)$, where
$\mathcal{B}(\R^n)$ denotes the Borel algebra of subsets of $\R^n$.
The usual notion of observable is then recovered as the special case
of a projection-valued measure (PVM) on $\br$, which is nothing but
the spectral measure associated with a selfadjoint operator. Observables 
represented are called PVMs {\em sharp} observables, all other POVMs 
are referred to as {\em unsharp} observables. The
extreme case of a {\em trivial} observable arises when all the
effects in its range are {\em trivial}, that is, of the form
$E(X)=\lambda_X\id$; the statistics associated with trivial effects
and observables carries no information about the state.

\subsection{Measurement schemes}

Measurements are physical processes and as such they are 
subject to the laws of
physics. In quantum mechanics, a measurement performed on an
isolated object is described as an interaction between this object
system and an apparatus system, both being treated as quantum
systems. Being a macroscopic system, the apparatus will interact
with a wider environment, but it is often convenient and sufficient
to subsume the degrees of freedom of this ``rest of the world" into
the description of the apparatus.

The quantum description of a measurement is succinctly summarized in
the notion of a {\em measurement scheme}, i.e., a quadruple
$\mathcal{M}:=\langle \mathcal{H}_\app,T_\app,U,Z\rangle$, where
$\mathcal{H}_\app$ is the Hilbert space of the apparatus (or probe)
system, $T_\app$ the initial apparatus state, $U=U(t_0,t_0+\Delta
t):\mathcal{H}\otimes\mathcal{H}_\app\to\mathcal{H}\otimes\mathcal{H}_\app$
is the unitary  operator representing the time evolution and ensuing
coupling between the object system and apparatus during the period
of measurement from time $t_0$ to $t_0+\Delta t$. Finally, $Z$ is
the apparatus pointer observable,  usually modeled as a sharp
observable.

\begin{figure}{
\includegraphics[width=0.9\textwidth]{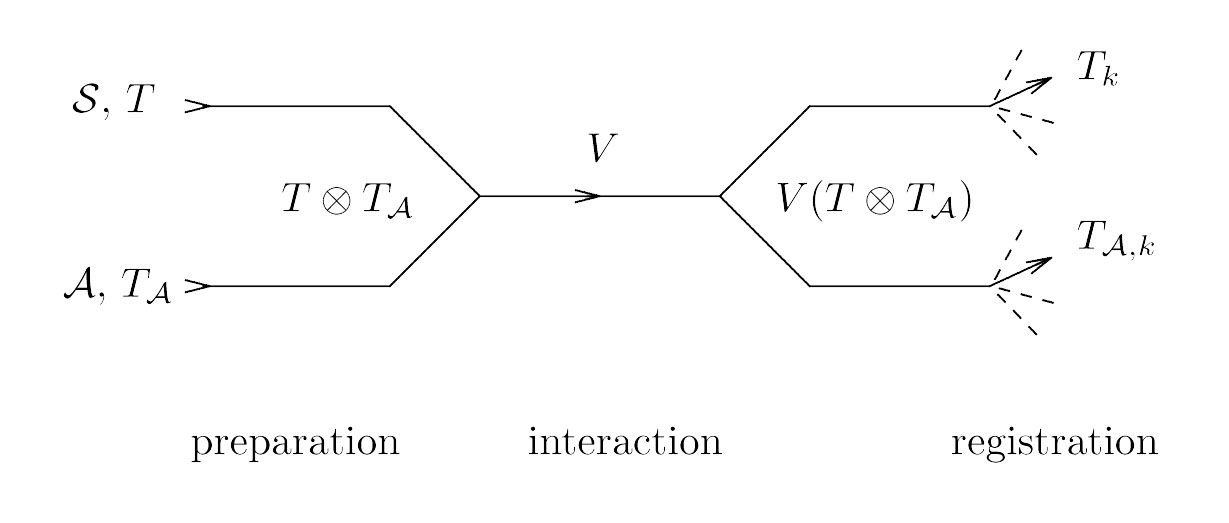} 
} \vspace{-0.8cm} \caption{Sketch of a measurement scheme. The symbols
are explained in the main body of the text.}\label{mmt-scheme}
\end{figure}

A schematic sketch of a measurement process is given in Figure
\ref{mmt-scheme} which is taken from  \cite{QTM}. Here $T$ and
$T_\app$ denote the initial states of the object and apparatus, and
$V(T\otimes T_\app):=UT\otimes T_\app U^*$ is the final state of the
compound system after the measurement coupling has ceased. It is
understood that upon reading an outcome, symbolized in the diagram
with a discrete label $k$, the apparatus is considered to be
describable in terms of a pointer eigenstate $T_{\app,k}$, and this
determines uniquely the associated final state $T_k$ of the object,
as will be shown below.

The observable measured by such a scheme is determined by the
pointer  statistics for every object input state and is thus
represented by a POVM $E$  that
is unambiguously defined by the following {\em probability reproducibility} condition:
\begin{equation}\label{eqn:prob-rep}
\tr{UT\otimes T_\app U^*\,I\otimes Z(X)}=:\tr{T\,E(X)}\equiv\prob^E_T(X).
\end{equation}
Here $X$ is any element of a $\sigma$-algebra $\Sigma$ of subsets of
an  outcome space $\Omega$. The positivity of the operators $E(X)$
in the range of the map $X\mapsto E(X)$ and the measure properties
of this map follows from the fact that the maps $X\mapsto
\prob^E_T(X)$ are probability measures for every state $T$.

The state $T_X$ of the object after recording a measurement outcome
in the set $X$ is determined by the following {\em sequential joint}
probability for a value of the pointer to be found in $X$ and an
immediately subsequent measurement of an effect $B$ to yield a
positive outcome:
\begin{equation}\label{eqn:mmt-instrument}
\tr{UT\otimes T_\app U^*\,B\otimes Z(X)}=:\tr{\stfm_X(T)\,B}\equiv\tr{T_XB}
\end{equation}
The maps $T\mapsto\stfm_X(T)=T_X$, called (quantum) operations, are
affine and trace norm-nonincreasing:
\begin{equation}\label{eqn:operations}
\tr{\stfm_X(T)}=\tr{T_X}= \tr{T\,E(X)}\leq \tr{T}=1,
\end{equation}
and they compose an {\em instrument}, that is, an operation-valued
map $X\mapsto\stfm_X$. Note that these maps $\stfm_X$ extend
in a unique way to linear maps on the complex vector space of trace
class operators. The above equation shows that every instrument
defines a unique POVM.

An important property of the operations $\stfm_X$ deriving from a
measurement scheme is their {\em complete positivity}: for every
$n\in\N$, the linear  map defined by $T\otimes\Theta\mapsto
\stfm_X(T)\otimes \Theta$ (where $T$ is any trace class operator on
$\hi$ and $\Theta$ is any trace class operator on $\C^n$) is
positive, that is, it sends state operators to (generally
non-normalized) state operators.\footnote{An example of a positive
state transformation that is {\em not} completely positive is given
by $T\mapsto CTC^*$, where $C$ is antilinear operator such as
complex conjugation $\psi(x)\mapsto\psi(x)^*$ for $\psi\in
L^2(\R)$.} The instrument composed of the completely positive
operations is also called completely positive.

Every measurement scheme defines thus a unique completely positive
instrument, and the latter fixes a unique POVM which represents the
observable measured by the scheme. Starting from
ground-breaking mathematical work of Neumark and Stinespring, the
converse statement was developed in increasing generality by Ludwig and
collaborators, Davies and Lewis, and Ozawa (detailed references can
be found in \cite{QTM}):
\begin{theorem}[Fundamental Theorem of Quantum Measurement Theory]\label{thm:fund-qtm}
\hfill\\
Every observable, represented as a POVM $E$, admits infinitely many completely positive
instruments $\stfm$ from which it arises via Eq.~(\ref{eqn:prob-rep}), and
every completely positive instrument admits infinitely
many implementations by means of a measurement scheme according to
Eq.~(\ref{eqn:mmt-instrument}).
\end{theorem}

\subsection{Examples}

Next I recall some model realizations of measurement schemes and completely
positive instruments; these will  provide valuable case
studies in subsequent sections.

\subsubsection{Von Neumann model of an unsharp position measurement}

On the final pages of his famous book of 1932, {\em  ``Mathematische Grundlagen der
Quantenmechanik"}, von Neumann introduces a mathematical model of  what he
describes as a measurement of the position of a particle in one spatial dimension.
Both the particle and measurement probe are represented by the Hilbert spaces
$\hi=\hi_\app=L^2(\R)$; and the coupling
\begin{equation}\label{eqn:vN-coupling}
U=\exp(-\tfrac i\hbar\lambda\qhat\otimes \phat_\app),
\end{equation}
generates a correlation between the observable intended to be measured, $\sfq$,
and the pointer observable $Z=\sfp_\app$.\footnote{The letters $\qhat,\phat$ denote
the selfadjoint canonical position and momentum operators, and their spectral measures
are denoted $\sfq,\sfp$, respectively.} 
To simplify the calculations,
one assumes that the interaction is {\em impulsive}, that is, the coupling constant
is large so that the duration of the interaction can be kept small enough so as to neglect
the free Hamiltonians of the two systems. It is further assumed that the initial state of the probe is
a pure state, $T_\app=\pee{\phi}$, with $\langle\qhat_\app\rangle_\phi:=\ip{\phi}{\qhat_\app\phi}=0$ and finite variance $\mathrm{Var}(\qhat_\app,\phi)=\langle\qhat_\app^2\rangle_\phi-
\langle\qhat_\app\rangle_\phi^2$.

Von Neumann proceeded to calculate the correlation between the particle's position
and the pointer observable {\em after} the coupling period and took this {\em measure
of repeatability} as an indication of the quality of the measurement. Had he made the
computation associated with equation (\ref{eqn:prob-rep}) above, he would have found
the actually measured observable to be a {\em smeared position} observable $\sfq_e$:\begin{equation}\label{eqn:smeared-pos}\begin{split}
E=\sfq_e:X\mapsto \sfq_e(X)&=\chi_X*e(\qhat)=\int_{\mathbb{R}}\chi_X*e(q)\,\sfq(dq),\\
\mathrm{where}\ e(q)&=\lambda|\phi(-\lambda q)|^2.
\end{split}\end{equation}
Here $*$ denotes the convolution.
Thus von Neumann was very close to discovering the representation of observables as
POVMs! The variance $\mathrm{Var}(\prob^{\sfq_e}_T)$ of the probability distribution
$\prob^{\sfq_e}_T$ is
\begin{equation}
\mathrm{Var}(\sfq_e,T)=\int_\R(x-\overline x)^2\prob^{\sfq_e}_T(dx)=
\mathrm{Var}(\sfq,T)^2+\mathrm{Var}(e),
\end{equation}
where $\overline{x}=\int_\R x\prob^{\sfq_e}_T(dx)=\tr{T\qhat}$. The second term in the
expression for the variance, $\mathrm{Var}(e)$, indicates the unsharpness of the observable
$\sfq_e$ and at the same time is a measure of the inaccuracy of the measurement, that is,
the separation between $\sfq$ and $\sfq_e$.

The instrument induced by von Neumann's measurement scheme is given as follows:
\begin{equation}\label{eqn:vN-instrum}\begin{split}
 \stfm^{\sfq_e}:\,X,T\mapsto\stfm^{\sfq_e}_X(T)&=\int_XK_qTK_q^*\,dq,\\
\mathrm{where}\  (K_q\fii)(x)&=\sqrt{\lambda}\phi\left(\lambda(q-x)\right)\fii(x).
\end{split}\end{equation}

\subsubsection{Ozawa's model of a sharp position measurement}

It turned out much more intricate to find a measurement scheme realizing a measurement of
the sharp position observable. One solution was presented by Ozawa \cite{Ozawa88,Ozawa02} who introduced the following coupling:
\begin{equation}\label{eqn:Ozawa-coupling}\begin{split}
U&=\exp\left[-\tfrac{i\pi}{3\sqrt 3\hbar}(2\qhat\otimes\phat_\app-2\phat\otimes\qhat_\app
+\qhat\phat-\qhat_\app \phat_\app) \right]\\
&=\exp\left(-\tfrac i\hbar\qhat\otimes\phat_\app\right)\,
\exp\left(\tfrac i\hbar\phat\otimes\qhat_\app\right).
\end{split}\end{equation}
Taking the pointer as $Z=\sfq_\app$, the measured observable is $\sfq$,
the sharp position, independently of the choice of initial probe state $T_\app$. Indeed,
the associated instrument is found to be
\begin{equation}\label{eqn:Ozawa-instr}
 \stfm^{\mathrm{Ozawa}}_X(T)=\int_X\tr{T\sfq(dq)}\,
e^{-\frac i\hbar q\phat}\,T_\app\,e^{\frac i\hbar q\phat},
\end{equation}
so that $\tr{TE(X)}=\tr{\stfm^{\mathrm{Ozawa}}_X(T)}=\tr{T\sfq(X)}$ for all states $T$ of the system.

\section{Quantum Limitations on Measurability}\label{sec:qu-lim}

The formalism of quantum measurements reviewed above provides a
framework for the rigorous formulation of limitations on the
measurability of physical quantities arising from quantum
structures.

\subsection{``No Information Gain Without Disturbance"}

There has been much debate over the claim that according to quantum
theory, every measurement necessarily ``disturbs" the object system.
Here is a theorem that states a precise sense in which this claim is
true.
\begin{theorem}
There is no instrument that leaves unchanged all states of the
system unless the associated observable is trivial. More precisely:
if an instrument $\stfm$ on $(\Omega,\Sigma)$ satisfies
$\stfm_\Omega(T)=T$ for all $T\in\sh$, then $T\mapsto
\tr{\stfm_X(T)}=:\lambda(X)$ is a constant map for all $X\in\Sigma$,
and so the induced observable $E$ is trivial, $E(X)=\lambda(X)\id$.
\end{theorem}

The proof is quickly sketched: if
$T=\pee{\fii}\mapsto\stfm_\Omega(\pee{\fii})=\stfm_X
(\pee{\fii})+\stfm_{\Omega\setminus X}(\pee{\fii})=\pee{\fii}$,
then $\stfm_X(\pee{\fii})=\lambda(X)\pee{\fii}$. Due to the
linearity of $\stfm_X$, the term $\lambda(X)$ is independent of
$\fii$, and the measured observable $E$ gives probabilities
independent of $\fii$:
$\prob^E_\fii(X)=\tr{\stfm_X(\pee{\fii})}=\lambda(X)$. {\em QED}

Hence a measurement scheme with no state change yields no
information gain. We note that ``disturbance" has here been
interpreted as state change. This conclusion immediately leads to
another question: is it possible to restrict the quality or accuracy
of a measurement and thereby control the extent of the disturbance?
This will be addressed in Section \ref{sec:inacc-dist}.

\subsection{``No measurement without (some transient) entanglement"}

It is a general fact of quantum mechanics that interactions between
two systems lead to {\em entanglement} between them, that is, to
states which are not of product form. From this it would seem to
follow that in a measurement the object system and apparatus end up
necessarily in an entangled state at the end of the coupling period.
The next theorem shows that this implication does not hold true
without qualifications.
\begin{theorem}\label{thm:mmt-entang}
Let $U:\mathcal{H}_{1}\otimes \mathcal{H}_{2}\rightarrow
\mathcal{H}_{1}\otimes \mathcal{H}_{2}$ be a \emph{non-entangling}
unitary measurement coupling such that for a fixed vector $\phi_0$
and all vectors $\varphi \in \mathcal{H}_{1}$, one has
$U(\varphi\otimes \phi )=\varphi ^{\prime }\otimes \phi ^{\prime }$.
Then $U$ acts in the following way:\newline (a)
$U(\varphi\otimes\phi_0)=V(\varphi)\otimes\phi'$, where $V$ is an
isometry;\newline (b) $U\left( \varphi \otimes \phi_0 \right)
=\varphi' \otimes W_{12}\varphi $, where
$W_{12}:\mathcal{H}_{1}\rightarrow \mathcal{H}_{2}$ is a surjective
isometry and\ $\varphi'$ is a fixed vector in $\hi_1$.
\end{theorem}
The proof is given in \cite{Busch03b}. From this result it follows
that if one aims at constructing a measurement scheme that leaves
the object and apparatus in a non-entangled (separable) state after
the coupling, and if this measurement is to transfer information
about the initial object state $\varphi$ to the apparatus, then the
coupling $U$ must act as in (b). It is therefore conceivable that
after a suitable coupling interaction has been applied, the object and
apparatus are left in an non-entangled state and yet complete
information about the object state has been transferred to the
apparatus. However, due to the continuity of the unitary dynamical
evolution $t\mapsto U_t$ which comprises the coupling operator
$U_{t+\Delta t}$, not all $U_{t'}$  with $t<t'<t+\Delta t$ can be of
the non-entangling form (b), since that operator is not continuously
connected with the identity operator $U_0$ at $t=0$. It follows that
that some intermittent entanglement must build up during the
interval $[t,t+\Delta t]$.

In order to extend this proof to measurement schemes for which the
initial apparatus state$T_\app$ is not pure, it is necessary to
sharpen the no-entanglement condition of the theorem to hold for any
vector in $\hi_\app$ whose projection operator can arise as a convex
component of $T_\app$. These vectors are known to be given exactly
by those in the range of $T_\app^{1/2}$ \cite{Hadjisavvas81}. The
following theorem, also proven in \cite{Busch03b}, can then be
applied to take a step towards extending the above discussion to mixed apparatus states.
\begin{theorem}
Let $U:\mathcal{H}_{1}\otimes \mathcal{H}_{2}\rightarrow
\mathcal{H}_{1}\otimes \mathcal{H}_{2}$ be a unitary mapping such
that for all vectors $\varphi \in \mathcal{H}_{1}$, $\phi \in
\mathcal{H}_{2}$, the image of $\mathcal{H}_{1}\otimes
\mathcal{H}_{2}$ under $U$ is of the form $U(\varphi\otimes \phi
)=\varphi ^{\prime }\otimes \phi ^{\prime }$. Then $U$ is one of the
following:\newline (A) $U=V\otimes W$ where
$V:\mathcal{H}_{1}\rightarrow \mathcal{H}_{1}$ and
$W:\mathcal{H}_{2}\rightarrow \mathcal{H}_{2}$ are unitary;\newline
(B) $U\left( \varphi \otimes \phi \right) =V_{21}\phi \otimes
W_{12}\varphi $, where $V_{21}:\mathcal{H}_{2}\rightarrow
\mathcal{H}_{1}$ and $W_{12}:\mathcal{H}_{1}\rightarrow
\mathcal{H}_{2}$ are surjective isometries.
\newline
The latter case can only occur if $\mathcal{H}_{1}$ and $\mathcal{H}_{2}$
are Hilbert spaces of equal dimensions.
\end{theorem}
It is not hard to construct a measurement scheme with a
non-entangling coupling of the form (B) for {\em any} object
observable $E$. This can be achieved by making the object interact
with another system of the same type onto which the state of the
original system is identically copied.

\begin{example} Let $\mathcal{H}_{1}=\mathcal{H}_{2}=\mathcal{H}$. Let
$E:\Sigma \rightarrow\eh $ be a POVM in $\mathcal{H}$. Define
$U\left( \varphi \otimes \phi\right) =\phi \otimes \varphi $. Then
we have
\begin{equation}
\langle U\varphi \otimes \phi |I\otimes E\left( X\right) U\varphi \otimes
\phi \rangle =\langle \varphi |E\left( X\right) \varphi \rangle .
\end{equation}
\end{example}

\subsection{``No repeatable measurement for continuous observables"}

\subsubsection{Repeatability and ideality}

A measurement and its associated instrument are called {\em repeatable} if the probability for
obtaining the same result upon immediate repetition of the measurement is equal to one:
\begin{equation}\label{eqn:rep}
\tr{\stfm_X\left(\stfm_X(T)\right)}=\tr{\stfm_X(T)}\quad \mathrm{for\ all}\ X\in\Sigma,\ T\in\sh.
\end{equation}

A measurement of a discrete observable and its associated instrument is called
{\em ideal} if it does not change any eigenstate; thus, if the state $T$ is such that
a particular outcome is certain to occur, then an ideal instrument does not alter
the state:
\begin{equation}\label{eqn:ideality}
 \mathrm{for\ all\ }T,k,\quad\mathrm{if\ }\tr{TP_k}=1\  \mathrm{then\ }\stfm_k(T)=T.
\end{equation}

Examples of repeatable measurements are the von Neumann  and L\"uders
measurements which will be defined next.

Let $A$ be an observable with discrete spectrum and associated spectral
decomposition $A=\sum_k a_kP_k$. We allow the eigenvalues to have multiplicity
greater than one, so that the spectral projections can be decomposed into a
sum of orthogonal rank-1 projections:  $P_k=\sum_\ell\pee{\fii_{k\ell}}$.
Then a {\em von Neumann measurement} is a measurement whose associated
instrument has the form
\begin{equation}\label{eqn:vN-inst-discrete}
\stfm_k^{\mathrm{vN}}(T)=\sum_\ell \pee{\fii_{k\ell}}T\pee{\fii_{k\ell}}.
\end{equation}
A {\em L\"uders measurement} is a measurement whose associated instrument
is given by:
\begin{equation}\label{eqn:Luders-instrum}
\stfm_k^{\mathrm{L}}(T)=P_kTP_k.
\end{equation}

Note that L\"uders measurement are ideal but von Neumann measurements are
not ideal if at least one eigenvalue is degenerate. The ideal measurements are
uniquely characterized by the form of their instruments \cite{QTM}:
\begin{theorem}\label{thm:ideal-mmt}
Any ideal measurement of a discrete sharp observable is a L\"uders measurement.
\end{theorem}
\noindent
In particular, it follows that every ideal measurement is repeatable. A much deeper
result is the following,
conjectured by Davies and Lewis  in 1970 \cite{DaLe70} and proven by M.~Ozawa
in 1984 \cite{Ozawa84}. An observable $E$ on $(\Omega,\Sigma)$ is {\em discrete}
if there is a countable subset of $N$ of $\Omega$ such that $E(N)=\id$.
\begin{theorem}\label{thm:rep-disc}
If a measurement of an observable $E$ is repeatable then $E$ is discrete.
\end{theorem}

I discuss briefly the implications of these results. First observe that the existence of ideal
measurements enables the applicability of the famous reality criterion of Einstein,
Podolsky and Rosen \cite{EiPoRo35}:
\begin{quote}
{\small ``If, without in any way disturbing a system, we can predict with certainty
(i.e., with probability equal to unity) the value of a physical quantity, then there
exists an element of physical reality corresponding to that physical quantity."}
\end{quote}
Since ideal measurements are repeatable, the associated observables must be
discrete. Hence the EPR criterion can only be applied to discrete observables
or discrete coarse-grainings of continuous observables.

\subsubsection{Approximate repeatability}

While strict repeatability is impossible for continuous observables such as position 
(or momentum), there do exist instruments for position (say) that are approximately
repeatable in the following sense. Let $\delta>0$, and for any (Borel) subset $X$ of 
$\R$ let $X_\delta$ denote the set of all points which have a distance of not more than
$\delta$ from some point in $X$. (Since $X_\delta=\bigcup_{x\in X}[x-\delta,x+\delta]$, this
set $X_\delta$ is a Borel set.) An instrument $\stfm$ on $\br$ is {\em $\delta$-repeatable}
if for all states $T$ and all $X\in\br$,
\begin{equation}
\tr{\stfm_{X_\delta}(\stfm_X(T))}=\tr{\stfm_X(T)}.
\end{equation}
An example is given by Ozawa's instrument of a sharp position measurement, 
Eq.~(\ref{eqn:Ozawa-instr}) if the probe state $T_\app$ is chosen such that its 
position distribution $\prob_{T_\app}^{\sfq_\app}$ is concentrated within $[-\delta,\delta]$.

The same form of instrument can also be defined for an unsharp position observable $\sfq_e$,
\begin{equation}
\stfm_X^{\sfq_e}(T)=\int_X\tr{T\sfq_e(dq)}e^{-iq\phat}T_\app e^{iq\phat},
\end{equation}
and if $T$ is chosen as before, one can find $d>0$ such that 
\begin{equation}
\tr{\stfm_{X_d}^{\sfq_e}(\stfm_X^{\sfq_e}(T))}\ge (1-\veps)\tr{\stfm_X^{\sfq_e}(T)}.
\end{equation}
Instruments with this property can be called $(d,1-\veps)$-repeatable.
A detailed proof can be found in \cite{BuLa90b}, and connections with the intrinsic
unsharpness of the observable $\sfq_e$ have recently been studied in \cite{CaHeTo06}.

\subsubsection{Approximate ideality}

Ideality is a form of nondisturbance, but it is restricted to the eigenstates of the
measured observable: if the quantity being measured has a definite value,
then such measurements do not change the state.  But any state other than an
eigenstate will be disturbed: it will be transformed into one of the eigenstates
due to the repeatability property of an ideal measurement.

The tight link between ideality and repeatability is relaxed if unsharp observables are
considered: these still allow a notion of approximate ideality, but that does not imply
approximate repeatability.  I illustrate the last statement by means of the {\em generalized L\"uders
instrument} associated with a discrete observable  $E:\omega_i\mapsto E_i$:
\begin{equation}\label{eqn:gen-Lud}
\stfm^{\mathrm{L}}_i(T)=E_i^{1/2}TE_i^{1/2}.
\end{equation}
The operations $\stfm^{\mathrm{L}}_i$ have the following  property:
\begin{equation}\label{eqn:approx-id}
\mathrm{ if}\ \tr{TE_i}\ge 1-\varepsilon\ \mathrm{ then\ }
\tr{\stfm^{\mathrm{L}}_i(T)E_i}\ge(1-\varepsilon)\tr{TE_i}.
\end{equation}
That is, they do not decrease the probability. Further, it can be shown that for
all states  $T$ for which $\tr{TE_i}\ge 1-\varepsilon$, the (trace norm) difference
between the states $T$ and $\stfm^{\mathrm{L}}_i(T)$ is of the order
$\varepsilon^{1/2}$; this is the sense in which the generalized L\"uders
instruments are approximately ideal. Approximately ideal measurements
enable a weakening of the EPR criterion applicable to unsharp or continuous
observables, thus yielding a notion of {\em unsharp reality} \cite{Busch85a}.

It is not hard to construct examples of effects (with some eigenvalues small)
such that the associated L\"uders operation does not increase the small
probability represented by that eigenvalue since the corresponding eigenstate
is left unchanged. This shows that repeatability does not hold even in an
approximate sense. Thus unsharp observables sometimes admit measurements
that are less invasive than measurements of sharp observables.

The notion of a L\"uders measurement was introduced by G. L\"uders in 1951
\cite{Luders51} (english translation in \cite{Luders51a}) who showed that such
measurements can be used to test the compatibility of sharp observables.
\begin{theorem}[L\"uders Theorem]\label{thm:Luders}
Let $A=\sum_k a_kP_k$ and $B$ be two (discrete) observable. The following are
equivalent:\\
(a) for all states $T$, $\tr{\sum_kP_kBP_k}=\tr{TB}$;\\
(b) $AB=BA$.
\end{theorem}
The statement also holds if the observable $B$ is not discrete or bounded; in that
case statements  (a) and (b) can be rephrased by replacing $B$ with all spectral
projections of $B$. This theorem has been used in relativistic quantum theory to
motivate the ``local commutativity" condition by virtue of the postulate that
measurements in one spacetime region should not lead to observable effects in
another, spacelike separated region.

According to the L\"uders theorem, any observable $B$ not commuting with $A$
is sensitive to a L\"uders measurement being performed on $A$. In other words,
a L\"uders measurement of $A$ disturbs the distributions of $B$ in some states
if $B$ does not commute with $A$. If $A,B$ are allowed to be unsharp observables,
the corresponding statement is no longer true in general but requires stronger
assumptions \cite{BuSi98}.
\begin{theorem}\label{thm:gen-Luders}
Let $E:\omega_i\mapsto E_i$ be a discrete observable and $B$ an effect. The
following are equivalent
if one of the assumptions (I) or (II) or (III) stated below holds:\\
(a') for all states $T$, $\tr{\sum_kE_k^{1/2}BE_k^{1/2}}=\tr{TB}$;\\
(b') $E_kB=BE_k$ for all $k$.\\
The assumptions are:\\
(I) $E$ is a simple observable with only two effects $E_1,E_2=\id-E_1$.\\
(II) $B$ has a discrete spectrum of eigenvalues that can be numbered in
decreasing or increasing order.\\
(III) Condition (a') is also stipulated for the effect $B^2$.
\end{theorem}
That {\em some} additional assumptions are necessary has been demonstrated
by means of a counter example in \cite{ArGhGu02}. There a discrete
unsharp observable $E$ and effect $B$ not commuting with $E$ were
found such that the generalized L\"uders instrument of $E$ does not disturb
the statistics of $B$.

\subsection{Measurement limitations due to conservation laws}

There is an obvious limitation on measurability due to the fact that
the physical realization of a measurement scheme depends on the
interactions available in nature. In particular, the Hamiltonian of
any physical system has to satisfy the symmetry requirements
associated with the fundamental conservation laws. This measurement
limitation is reviewed in Abner Shimony's contribution, so that here
some complementary points and comments will be sufficient.

An early demonstration of the impact of the existence of additive
conserved quantities on the measurability of a physical quantity was
given by Wigner in 1952 \cite{Wigner52}. Wigner showed that
repeatable measurements of the $x$-component of a spin-1/2 system
are impossible due to the conservation of the $z$-component of the
total angular momentum of the system and the apparatus. The
conclusion was generalized by other authors to the statement that a
repeatable measurement of a discrete quantity is impossible if there
is a (bounded) additive conserved quantity  of the object plus
apparatus system that does not commute with the quantity to be
measured.

Wigner's resolution was to show that a successful measurement can be
realized with an angular-momentum-conserving interaction and
with an arbitrarily high success probability if the
apparatus is sufficiently large. Thus he allowed for an additional
measurement ``outcome" that indicated ``no information" about the
spin. The outcomes associated with ``spin up" and ``spin down" were
shown to be reproduced with probabilities that came arbitrarily
closely to the ideal quantum mechanical probabilities. In \cite[Sec.
IV.3]{OQP} it was shown that this resolution amounts to describing
the measurement by means of a POVM with three possible outcomes and
associated effects $E_+,E_-,E_?$, where the effects
$E_\pm=(1-\varepsilon)P^{s_x}_\pm$, i.e., they are ``close to" the
spectral projections of $s_x$ if $0<\varepsilon\le 1$, and the
effect $E_?=\varepsilon\id$ is a multiple of $\id$. It can be shown
that $\varepsilon$ can be made very small if the size of the measuring
system is large.

These considerations show that it is a matter of principle that measurements
of spin can never be perfectly accurate as a consequence of
the additive conservation law for total angular momentum.
The the necessary inaccuracy is appropriately described
by a POVM of the kind described above. However, the common description
of a sharp spin measurement is found to be an admissible idealization; the
error made by breaking (ignoring) the fundamental rotation symmetry of the
measurement Hamiltonian is negligible due to the fact that the measuring
system is very large.

It seems to be a difficult problem to decide whether a limitation of measurability
arises also in cases where the observable to be measured and the conserved
quantity are unbounded and have  continuous spectra. This question was raised
by Shimony and Stein in 1979 \cite{ShSt79}. The most general result at that time
was the following (expressed in the notation of the present paper):
\begin{theorem}\label{thm:mmt-cons-law}
If a sharp observable $E$ admits a repeatable measurement, and if
$L\otimes \id + \id\otimes L_\app$ is a bounded selfadjoint operator representing a
conserved quantity for the combined object and apparatus system, then
$E$ commutes with $L$.
\end{theorem}
Since repeatable measurements exist only for discrete observables
(Theorem \ref{thm:rep-disc}), the above statement is only applicable
to object observables with  discrete spectra. Hence it does not
apply to measurements of position.

Ozawa \cite{Ozawa91}
presented what seems to be a counter example, using a coupling that is manifestly translation
invariant. However, this model constitutes an unsharp position measurement which becomes a sharp measurement only if the initial state of the apparatus is allowed to be a non-normalizable state (that is,
not a Hilbert space vector or state operator).\footnote{The same observation applies to the von Neumann measurement model of which Ozawa's model is a modification.} A proof that a sharp position measurement (without repeatability, but with some additional physically reasonable assumptions)
cannot be reconciled with momentum conservation
was given in \cite{Busch85b}. A general proof is still outstanding.

Here we use another modification of the von Neumann model to demonstrate that
momentum conservation is compatible with unsharp position measurements where
the inaccuracy can be made arbitrarily small \cite[Sec. 4.3]{OQP}.
Note that the total momentum $\phat+\phat_\app$ commutes with the coupling
\begin{equation}\label{eqn:mod-vN-coupling}
U=\exp\left(-i\tfrac \lambda 2\bigl[(\qhat-\qhat_\app)\phat_\app +
\phat_\app(\qhat-\qhat_\app)\bigr]\right).
\end{equation}
The pointer is again taken to be $Z=\sfq_\app$. Then the
measured observable is the smeared position $\sfq_e=e*\sfq$, where
$e(q)=\left(e^{\lambda}-1\right)\,\Bigl|\phi\bigl(-(e^{\lambda}-1)q\bigr)\Bigr|^2 $.

One can argue that the clash between the conservation law and
position measurement has been shifted and reappears when the
measurement of $\sfq_\app$ is considered. However, if momentum
conservation is taken into account in the measurement of the
pointer, it would turn out that the pointer itself is only measured
approximately,  that is, an unsharp pointer $\sfq_{\app,h}$ is
actually measured,  which then yields the measured observable as
$\sfq_{e*h}$.

The lesson of the current subsection is this: to the extent  that
the limitation on measurability due to additive conservation laws
holds as a general theorem, it shows that the notion of a sharp
measurement of the most important quantum observables
is an idealization which can be realized only approximately {\em as
a matter of principle}; yet the quality of the approximation can be
extremely good due to the macroscopic nature of the measuring
apparatus.

To conclude this section, it is worth remarking that the quantum limitations of
measurements described here are valid independently of the view that one
may take on the measurement problem. This is the case because these limitations
follow from consideration of the total state of system and apparatus as it arises
in the course of its unitary evolution. 

\section{Complementarity and Uncertainty}\label{sec:CU}

The {\em ``classic"} expressions of quantum limitations of preparations
and measurements are codified in the complementarity and uncertainty
principles, formulated by Bohr and Heisenberg 80 years ago.

This section offers a ``taster" for two recent extensive reviews on
the complementarity principle, Ref.~\cite{BuSh06}, and the uncertainty
principle, Ref.~\cite{BuHeLa06}, which together develop a novel coherent
account of these two principles. 
In a nutshell, complementarity states a strict exclusion of certain pairs of operations
whereas the uncertainty principle shows a way of ``softening" complementarity into
a graded, quantitative relationship, in the form of a trade-off between the accuracies
with which these two options can be realized together approximately. This interpretation
is compatible with, if not envisaged in, the following passage of Bohr's published
text of his famous Como lecture of 1927 \cite{Bohr28}.
\begin{quote}
{\small ``In the language of the relativity theory, the content of the relations (2) [the
uncertainty relations] may be summarized in the statement that according to the
quantum theory a general reciprocal relation exists between the maximum
sharpness of definition of the space-time and energy-momentum vectors
associated with the individuals. This circumstance may be regarded as a simple
symbolical expression for the complementary nature of the space-time
description and claims of causality. At the same time, however, the general
character of this relation makes it possible to a certain extent to reconcile
the conservation laws with the space-time co-ordination of observations, the
idea of a coincidence of well-defined events in a space-time point being
replaced by that of \emph{unsharply} defined individuals within finite space-time
regions."}
\end{quote}
Bohr summarizes here his idea of complementarity as the falling-apart in quantum
physics of the notions of observation, which leads to {\em space-time description},
and state definition,  linked with {\em conservation laws} and {\em causal description};
he regarded the possibility of combining space-time description and causal
description as an idealization that was admissible in classical physics. Note also
the reference to {\em unsharpness} (the emphasis in the quotation is ours), which
seems to constitute the first formulation of an intuitive notion of {\em unsharp reality}
(and the first occurrence of this teutonic addition to the English language).

\subsection{The Complementarity Principle}

In a widely accepted formulation, the {\em Complementarity
Principle} is the statement that there are pairs of observables
which stand in the relationship of complementarity. That
relationship comes in two variants, stating the mutual exclusivity
of {\em preparations} or {\em measurements} of certain pairs of
observables. In quantum mechanics there are pairs of observables the
eigenvector basis systems of which are mutually unbiased. This means
that the system is in an eigenstate of one observable, so that the
value of that observable can be predicted with certainty, the values
of the other observable are uniformly distributed. This feature is
an instance of {\em preparation complementarity}, and it has been
called {\em value complementarity}. {\em Measurement
complementarity} of observables with mutually unbiased eigenbases
can be characterized by the following property: any attempt to
obtain simultaneous information about both observables by first
measuring one and then the other is bound to fail since the first
measurement completely destroys any information about the other
observable; that is to say, the second measurement gives no
information about the state prior to the first measurement. This
will be illustrated in an example below. We conclude that the
``principle" of complementarity, as formalized here, is in fact a
consequence of the quantum mechanical formalism.

Examples of pairs of observables are spin-1/2 observables such as
$s_x,s_z$, and the canonically conjugate position and momentum
observables $\qhat,\phat$ of a free particle. A unified
formalization of preparation and measurement complementarity can be
given in terms of the spectral projections of these observables
($P^x_\pm,P^z_\pm$ for $s_x,s_z$, and $\sfq(X),\sfp(Y)$ for
$\qhat,\phat$:
\begin{equation}\label{eqn:complem}\begin{split}
P^x_k\land P^z_\ell&=\nul\quad \mathrm{for\ } k,\ell=+,-\,;\\
\sfq(X)\land\sfp(Y)&=\nul \quad  \mathrm{for\ bounded\ intervals}\ X,Y.
\end{split}\end{equation}
The symbol $\land$ represents the lattice-theoretic infimum of two
projections,  that is, for example, $\sfq(X)\land\sfp(Y)$ is the
projection onto the closed subspace which is the intersection of the
ranges of $\sfq(X)$ and $\sfp(Y)$. These relations entail, in
particular, that complementary pairs of observables do not possess
joint probability distributions associated with a state $T$ in the
usual way: for example, there is no POVM $G:\brr\to\eh$ such that
$G(X\times\R)=\sfq(X)$ and $G(\R\times Y)=\sfp(Y)$ for all
$X,Y\in\br$. In fact, if these marginality relations were satisfied
for all bounded intervals $X,Y$, then one must have $G(X\times Y)\le
\sfq(X)$ and $G(Y\times Y)\leq \sfp(Y)$, and this implies that any
vector in the range of $G(X\times Y)$ must also be in the ranges of
$\sfq(X)$ and $\sfp(Y)$, hence $G(X\times Y)=\nul$.

\begin{example}[Complementarity for measurement sequences (1)] Let
$A,B$ be observables in $\C^n$, $n\ge 2$, with mutually unbiased
eigenbases $\fii_1,\fii_2,\dots,\fii_n$ and
$\psi_1,\psi_2,\dots,\psi_n$, respectively. (Hence $A,B$ are value complementary.)
Let $\stfm^A$ be the repeatable (von
Neumann-L\"uders) instrument associated with  $A$:
$\stfm_k^A(T):=\ip{\fii_k}{T\fii_k}\kb{\fii_k}{\fii_k}$. Let
$\stfm_\R^A:=\sum_k\stfm^A_k$ be the nonselective measurement
operation, then the probability for a $B$ measurement following the
$A$ measurement is $\prob^B_{\stfm_\R(T)}(\ell)=1/n$, which is
independent of $T$. This can be expressed by saying that the
observable effectively measured in this process is not $B$ but the
trivial POVM whose effects are $E_\ell=\frac 1n\id$.
\end{example}

\begin{example}[Complementarity for measurement sequences (2)]
Consider a measurement of position $\qhat$ followed by a measurement
of momentum $\phat$. Let $\stfm^\sfq$ be the  instrument
representing the position measurement. Then the following defines a
joint probability distribution:
\begin{equation}
\tr{\stfm^\sfq_X(T)\,\sfp(Y)}=\prob_T(X\times Y)=:\tr{TG(X\otimes Y)},\quad X,Y\in\br.
\end{equation}
The marginal observables are sharp position and a ``distorted
momentum" observable, $G(X\times \R)=\sfq(X)$ and $G(\R\times
Y)=\widetilde\sfp(Y)$. Since one of these marginal observables is a
sharp observable, it follows that the effects of the other marginal
observable commute with the sharp observable. But $\sfq$ is a
maximal observable, and so the effects $\widetilde\sfq(Y)$ are in
fact functions of the position operator. The attempted momentum
measurement only defines an effectively measured observable which
contains a ``shadow" of the information of the first position
measurement. Hence a sharp measurement of position destroys all
prior information about momentum (and vice versa).

The following defines a completely positive instrument $\stfm^\sfq$
which renders the effective observable defined by a subsequent
momentum measurement trivial: let $T_x$ be the continuous family of
positive operators  of trace one, generated by
$T_x:=U_xT_0U_x^{-1}$, where $U_x$ are unitary operators that
commute with momentum $\phat$. Then put
\begin{equation}\label{eqn:Q-rep-inst}
\stfm^\sfq_X(T):=\int_XT_x\,\tr{T\sfq(dx)}.
\end{equation}
The associated measured observable is indeed the sharp observable $\sfq$ since
$\tr{\stfm^\sfq_X(T)}=\tr{T\sfq(X)}$. Then the distorted momentum observable
$\widetilde{\sfp}$ defined above is found to be:
\begin{equation}\begin{split}
\tr{T\widetilde{\sfp}(Y)}:=&\tr{\stfm^\sfq_\R(T)\sfp(Y)}=
\int_\R\tr{T_x\sfp(Y)}\tr{T\sfq(dx)}\\
=&\int_\R\tr{T_0U_x^{-1}\sfp(Y)U_x}\tr{T\sfq(dx)}\\
=&\int_\R\tr{T_0\sfp(Y)}\tr{T\sfq(dx)}=
\tr{T_0\sfp(Y)}.
\end{split}\end{equation}
Thus $\widetilde{\sfp}$ is a trivial observable. Note that in this
calculation  $\sfq$ could have been replaced by any observable as
the first-measured observable. However, if the instrument
(\ref{eqn:Q-rep-inst}) is required to be approximately repeatable, then
$T_0$ must have a position distribution concentrated around the
origin 0, and $U_x$ must ensure that $T_x$ has a position
distribution concentrated around the point $x$; this is achieved if
$U_x$ is chosen to be $exp(\frac i\hbar x\phat)$. Notice that this
form is in fact realized in the Ozawa instrument for a sharp
position measurement,Eq.~(\ref{eqn:Ozawa-instr}). While we have not
shown that this form is necessary, this consideration suggests that
for approximately repeatable position measurements a subsequent
momentum measurement leads to a (nearly) trivial observable as the
distorted momentum.
\end{example}


\subsection{The Uncertainty Principle}

Following Ref.~\cite{BuSh06}, we propose that the term \emph{uncertainty principle}
refers to the broad statement that there are  pairs of observables for which
a trade-off relationship pertains for the degrees of sharpness of the
preparation or measurement of their values, such that a simultaneous
or sequential determination of the values requires a nonzero amount of
unsharpness (latitude, inaccuracy, disturbance). This gives rise to {\em three} variants
of uncertainty relations, exemplified here for position and momentum:
first there is the well-known inequality for the widths of  the probability
distributions of position and momentum in any quantum state that can be expressed
in terms of the standard deviations,
\begin{equation}
\Delta(\sfq,T)\Delta(\sfp,T)\ge\tfrac12\hbar.
\end{equation}
Second, one may consider a trade-off relation for the inaccuracies in any attempted {\em joint measurement} of position and momentum,
\begin{equation}
\delta(\widetilde{\sfq},\sfq)\,\delta(\widetilde{\sfp},\sfp)\ge C\hbar,
\end{equation}
where the inaccuracies are to be defined appropriately as measures of the differences
between the sharp position and momentum observables $\sfq,\sfp$ and their approximations
$\widetilde{\sfq},\widetilde{\sfp}$, respectively, which are to be measured jointly. Finally,
there is a trade-off between the accuracy of an approximate measurement of position (momentum)
and a necessary disturbance of the momentum (position) distribution:
\begin{equation}
\delta(\widetilde{\sfq},\sfq)\, D(\widetilde\sfp,\sfp)\ge C\hbar,\quad
\delta(\widetilde{\sfp},\sfp)\, D(\widetilde\sfq,\sfq)\ge C\hbar,
\end{equation}
where $D(\widetilde\sfq,\sfq)$ and $D(\widetilde\sfp,\sfp)$ denote appropriate measures of the
disturbance of position and momentum, respectively.

Suitable measures of inaccuracy and disturbance which make the last two measurement uncertainty relations precise will be presented in Section \ref{sec:inacc-dist}. It thus turns out that similar to the complementarity principle, the uncertainty principle in its three manifestations is also a formal consequence of the noncommutativity of the observables in question. The term ``principle" may still be used to highlight the fact that the uncertainty relations reflect an important nonclassical feature of quantum mechanics.

\subsection{Complementarity versus uncertainty?}

The reviews \cite{BuSh06} and \cite{BuHeLa06} propose a resolution of a long-standing
controversy over the relationship, relative roles and interplay of the complementarity and
uncertainty principles. This resolution will be briefly summarized here.  As indicated in the
introductory quote from Bohr (1928), the traditional view describes the uncertainty relations
as a formal expression of the complementarity principle. However, as a quick survey of
the research and textbook literature on quantum mechanics shows, this view has met with a considerable degree of uneasiness by many. Some authors consistently avoid any reference to
complementarity while others play down the significance of the uncertainty relations, denying them
the status of a principle which they reserve for complementarity.

Yet, in recent years there has been a shift of perspective which was indeed anticipated in
the same quote of Bohr: complementarity is seen as a statement of the {\em impossibility}
of jointly performing certain pairs of preparation or measurement procedures, whereas the role of the uncertainty principle is to quantify the degree to which an approximate reconciliation of these
mutually exclusive options becomes a {\em possibility}. It seems that in this way a more balanced assessment has been achieved: compared to the view that emphasized complementarity over
uncertainty, the positive role of the uncertainty relations as enabling joint
determinations and joint measurements is now highlighted more prominently; and
even though it is true (as shown in \cite{BuSh06}) that the uncertainty relations
entail the complementarity relations in a suitable limit sense, it is still appropriate to point out
the strict mutual exclusivity of sharp value assignments which, after all, is
the reason for the quest for an approximate reconciliation in the form of
simultaneous but unsharp value assignments.

The principles of complementarity and uncertainty are extreme manifestations of the existence of noncommuting pairs of observables and of superpositions of states, which both entail fundamental limitations of the possibilities of preparing or measuring simultaneous sharp values of observables that do not commute. These limitations are consequences of a famous theorem of von Neumann which we summarize here as follows.
\begin{theorem}\label{thm:vN}
Let $A$ and $B$ be two sharp observables represented as selfadjoint operators. The following are equivalent:\\
(a) $A$ and $B$ possess a joint spectral representation (possibility of preparing joint sharp values).\\
(b) $A$ and $B$ possess a joint observable that defines joint probabilities for them
(jointly measurability).\\
(c) $AB=BA$.
\end{theorem}

The reason for the long-standing debate over the superiority of either the complementarity
{\em principle}  or the uncertainty {\em principle} seems to lie in the fact that the {\em features} of complementarity and uncertainty are formally intertwined in Hilbert space  quantum mechanics.
It is only in the context of theoretical frameworks more abstract and general than quantum or classical theories that the logical relationships between complementarity and uncertainty postulates can be
investigated; in such a generalized setting these postulates  can in fact be used as principles within
a set of axioms from which the Hilbert space framework of quantum mechanics can be deduced.
As an example, we note the work of P.~Lahti together with the late S.~Bugajski, Ref.~\cite{LaBu85},
who used appropriate formalizations of complementarity and the existence of von Neuman-L\"uders
measurements in the so-called convexity framework to derive Hilbert space quantum theory.

\section{Inaccuracy and disturbance in quantum measurements}\label{sec:inacc-dist}

It remains to show how the above programmatic statement of the uncertainty principle for joint and sequential measurements can be made precise by appropriate measures of inaccuracy and disturbance. Such measures are also applicable in the analysis of the other quantum limitations of measurability discussed in Sec.~\ref{sec:qu-lim}.

First I will introduce the idea of an approximate joint measurement of two noncommuting
quantities and present an operational definition of measurement error applicable to
continuous observables such as position and momentum; the error measures for these
observables obey a trade-off relation valid in any approximate joint measurements. Then
I will show that a trade-off relation between the accuracy of a measurement and the
disturbance of the distributions of an observable not commuting with the measured
observable can be considered as an instance of a trade-off relation between the
inaccuracies in an approximate joint measurement of two noncommuting observables.

\subsection{Approximate joint measurements}

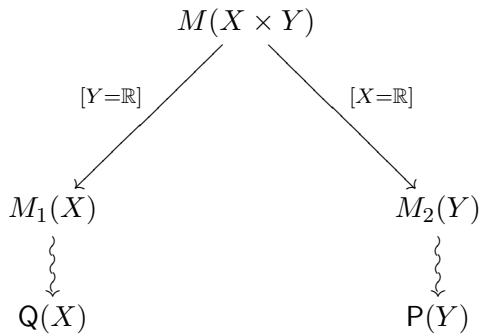
\begin{figure}
\[
\xymatrix{
& M(X\times Y)\ar[ldd]_{\small [Y=\mathbb R]}\ar[rdd]^{\small [X=\mathbb R]} &\\
&&\\
M_1(X)\ar@{~>}[d]&&M_2(Y)\ar@{~>}[d]\\
\sfq(X)&&\sfp(Y)}
\]
\caption{Idea of a joint approximate measurement of position $\sfq$ and momentum $\sfp$,
by means of an observable $M$ on $\brr$ whose marginals $M_1$ and $M_2$ are
approximations of $\sfq$ and $\sfp$, respectively.}\label{fig:approx-joint-meas}
\end{figure}

A necessary criterion for the joint measurability of two observables is the existence of a
joint probability distribution for every state $T$ in the usual quantum mechanical form.
Von Neumann's theorem entails that two noncommuting sharp observables  such as
position and momentum do not possess joint distributions (for all states). Hence these observables are not jointly measurable. However, for the joint measurability of pairs of {\em unsharp} observables, commutativity is {\em not} a necessary requirement. This
suggests the following consideration: it should be possible to find two jointly
measurable observables $M_1,M_2$ on $\br$ which are {\em approximations}, in a suitable
sense, of position $\sfq$ and momentum $\sfp$, respectively. Then a measurement of a joint
observable $M$ on $\brr$ of $M_1,M_2$ will be accepted as an {\em approximate joint
measurement} of $\sfq,\sfp$ if  the deviations of $M_1$ from $\sfq$ and of
$M_2$ from $\sfp$ are finite in some appropriate measure. This constellation is shown in
Figure \ref{fig:approx-joint-meas}.

Two tasks need to be addressed in order to complete the above program. First, one
needs to introduce suitable operational measures of inaccuracy, that is, of the deviation
between two observables defined on the same outcome space $(\Omega,\Sigma)$. Second,
since we are interested in good joint approximations of noncommuting pairs of observables,
the optimal approximators $M_1,M_2$ must be expected to be noncommuting and hence
unsharp observables in order to be jointly measurable; therefore, the problem arises to
quantify the necessary degree of unsharpness required for the joint measurability given the
finite ``distance" of $M_1,M_2$ from two noncommuting observables.

The definition of such measures of inaccuracy and unsharpness will in general depend
on the type of outcome space. A variety of approaches for the case $(\R,\br)$ are
analyzed in \cite{BuHeLa06} and compared in detail in \cite{BuPe07}, and the case of
discrete (qubit) observables is investigated in \cite{BuHe07}. Here I will give a brief survey
of notions applicable to the position-momentum case.

\subsubsection{Standard error}

The only known measure that is universally applicable to different types of outcome spaces
(barring questions of domains of unbounded operators) is a quantity that may be called
{\em standard error} as it is defined in terms of the first and second moments of the relevant
operator measures, similar to the standard deviation. This seems to be the only measure of
inaccuracy or error that has been in use in the literature over an extended period. Examples of
its application in the formulation of uncertainty relations for joint measurements are the
works of Appleby \cite{Appleby98a,Appleby98b}, Hall \cite{Hall04}, and Ozawa (e.g.,
\cite{Ozawa03a,Ozawa04b}).

For an observable $E$ on $\br$, let $E[k]:=\int x^kE(dx)$ denote the $k^{th}$ moment
operator of $E$ (defined on its natural domain
$D(E[k]):=\{\fii\in\hi\,:\,\left|\int x^k\ip{\psi}{E(dx)\fii}\right|<\infty\ \mathrm{for\ all\ }\psi\in\hi\}$
\cite{DvLaYl00}).
Assume $\mathcal{M}$ is a measurement scheme defining an observable $E$ on $\br$
which is intended to approximate the sharp position $\sfq$. Then a suggestive choice of
measure of inaccuracy is
\begin{equation}\label{eqn:stand-err01}
\varepsilon(Z,\sfq;T):=\tr{UT\otimes\pee{\phi}U^*(\id\otimes Z[1]-\qhat\otimes\id)^2}^{1/2}.
\end{equation}
This can be expressed in terms of the actually measured observable $E$:
\begin{equation}\label{eqn:stand-err02}
\varepsilon(E,\sfq;T):=\left(\tr{T(E[1]-\qhat)^2}+\tr{T(E[2]-E[1]^2)}\right)^{1/2}.
\end{equation}
The inaccuracy in a momentum measurement is defined similarly. Ozawa proved the following
universal uncertainty relation for the marginals $M_1,M_2$ of an observable $M$ on $\brr$:
\begin{equation}\label{eqn:stand-err-ur-Ozawa}
\varepsilon(M_1,\sfq)  \varepsilon(M_2,\sfp)+
\varepsilon(M_1,\sfq)\Delta(\sfp,T) +\Delta(\sfq,T)\varepsilon(\sfp,T)  \ge\frac12 \hbar .
\end{equation}
He noted that the first product term can be zero (this happens in Ozawa's model of a sharp position
measurement introduced above), and considers
this to be a demonstration that the Heisenberg uncertainty principle for joint measurements
of position and momentum and that for inaccuracy vs disturbance does not have the common
form with a state-independent lower bound.

However, this way of reasoning ignores two crucial deficiencies in the definition of
$\varepsilon(E,\sfq;T)$ as a measure of inaccuracy. First, the above uncertainty relation is not a
statement solely about measurement inaccuracies since it depends on the preparation of the
system. An appropriate definition of measurement inaccuracy should give an estimate of error
which can be obtained without reference to the state of the measured object (which usually is
unknown in a measurement). This point was observed by Appleby in 1998 who introduced what
we propose to call the {\em (global) standard error}:
\begin{equation}\label{eqn:glob-st-err}
\varepsilon(E,\sfq):=\sup_{T\in\sh}\left(\tr{T(E[1]-\qhat)^2}+\tr{T(E[2]-E[1]^2)}\right)^{1/2}.
\end{equation}
This quantity gives rise to a universal trade-off relation for joint measurement errors.
\begin{theorem}\label{thm:st-err-ur}
Let $M$ be an observable on $\brr$. Its marginals $M_1,M_2$ obey the following:
\begin{equation}\label{eqn:st-err-ur}
\varepsilon(M_1,\sfq)\varepsilon(M_2,\sfp)\ge\tfrac 12\hbar.
\end{equation}
\end{theorem}
\noindent
Appleby \cite{Appleby98b} gave a simple informal derivation. A rigorous proof is given in
\cite{BuPe07}.

The second deficiency of the definition of $\varepsilon(E,\sfq;T)$ -- and also of
$\varepsilon(E,\sfq)$ -- lies in the fact that this quantity
cannot be estimated in terms of the measurements of $E$ and $\sfq$ under consideration unless
the operators $E[1]$ and $\sfq$ commute so that they can be jointly measured to determine the
expectation of the operator $(E[1]-\sfq)^2$. If $E[1]$ and $\sfq$ do not commute then normally
the squared difference operator does not commute with either of them and a third, quite different
measurement is required to find its expectation value. This is to say that the standard error
is not {\em operationally significant}, in general.

An interesting but very special subclass of measurements where this deficiency does not
arise is the family of {\em unbiased} measurements, for which $E[1]=\sfq$. In this case the
standard error is given solely by the second term in Eq.~(\ref{eqn:stand-err02}), which is
actually an operational measure
of the intrinsic {\em noise} or unsharpness of the approximator $E$ of $\sfq$ (see below).

\subsubsection{A distance between observables on $\br$}

In 2004, R.~Werner \cite{Werner04b} introduced a
distance $d(E,F)$ between two observables $E$ and $F$ on $\br$ which is sensitive to the
distance of the bulks of probability distributions $\prob_T^E$ and $\prob_T^F$, and
he derived an uncertainty relation for position and momentum. Some definitions are required
in order to present this result.

For any bounded continuous function $g$ on $\R$, one can define the operator $L(g,E):=\int_\R g(x)E(dx)$.
The definition of $d(E,F)$ makes use of the set of (Lipshitz) functions
$\Lambda:=\{g:\mathbb{R}\to\mathbb{R}\,:\, g\ \mathrm{bounded},\ |g(x)-g(y)\le|x-y|\}$.
Werner's distance then is given as follows:
\begin{equation}\label{eqn:distance}
d(E,F):=\sup\big\{\,\|L(g,E)-L(g,F)\|\, :\, g\in\Lambda \big\}
\end{equation}
Werner's joint measurement uncertainty relation is stated as follows \cite{Werner04b}.
\begin{theorem}\label{thm:Werner}
Let $M_1,M_2$ be marginals of an observable $M$ on $\brr$. The distances $d(M_1,\sfq)$
and $d(M_2,\sfp)$ obey the inequality
\begin{equation}\label{eqn:Werner-ur}
d(M_1,\sfq)\,d(M_2,\sfp)\ge C\hbar.
\end{equation}
Here the optimal constant $C$ is determined via $C\hbar=E_0^2/(4ab)$, where $E_0$ is
the lowest (positive) eigenvalue of the operator $a|Q|+b|P|$ for some $a,b>0$. Its value is
given by $C\approx 0.304745$.
\end{theorem}
\noindent
This result constitutes the first universal joint measurement inaccuracy relation for
operationally significant measures of inaccuracy. Moreover, the proof techniques used turn
out to be applicable for quite different definitions of inaccuracy (see \cite{BuPe06,BuPe07}).
The distance $d(E,F)$ is geometrically appealing and constitutes a natural choice due to its
connection with the so-called Monge metric on the space of probability measures on $\br$.
However, from an experimenter's perspective, it may be considered less appealing to be
asked to estimate $d(E,F)$ by measuring differences of expectation values for $L(g,E)$
and $L(g,F)$, where $g$ runs through the set $\Lambda$ of Lipshitz functions.

\subsubsection{Error bar width}

A measure of measurement inaccuracy that would appear natural to an experimenter is the
width of error bars, which is estimated in a process of calibration: the measurement scheme
to be calibrated is fed with systems prepared with fairly sharply defined values of (say)
the position observable. For each value, one estimates the spread of output values which gives
a measure of the {\em error bar width}. If this measure is found to be bounded across all
input values, the measurement will be considered to constitute a good approximation of
the position observable to be measured. This consideration is captured in the following definitions.

Let $M_1$ be an observable on $\br$ which is to approximate $\sfq$. Let $\iqd:=
[q-\delta/2,q+\delta/2]$.
By $\delq{M_1}$ I denote the {\em inaccuracy}, defined as the smallest interval width $w$ such
that whenever the value of $\sfq$ is certain to lie within an interval $\iqd$, then the output
distribution $\prob_\fii^{M_1}$ is concentrated to within $1-\veps_1$ in $\iqw$:
\begin{equation}\label{eqn:inacc}\begin{split}
\delq{M_1}:=\inf\{w\,|\ &\text{for all\ }q\in\R,\,\psi\in\hi,\\
& \text{if\ } \prob^\sfq_\psi(J_{q;\delta})=1
\ \text{then\ }\prob_\psi^{M_1}(J_{q;w})\ge 1-\veps_1\}.
\end{split}\end{equation}

The inaccuracy describes the range within which the input values can be inferred
from the output distributions, with confidence level $1-\varepsilon$, given initial localizations within
$\delta$.
The inaccuracy is an increasing function of $\delta$, so that one can
define the \emph{error bar width} of $M_1$ relative to $\sfq$:
\begin{equation}\label{error-bar}
\deq{M_1}:=\inf_\delta\delq{M_1}=\lim_{\delta\to 0}\delq{M_1}.
\end{equation}
If $\deq{M_1}$ is finite for all $\veps_1\in (0,\frac 12)$, we will say that $M_1$ approximates
$\sfq$ in the sense of \emph{finite error bars}.
Similar definitions apply to approximations $M_2$ of momentum $\sfp$, yielding
$\delp{M_2}$ and $\dep{M_2}$.

It is interesting to note that  the finiteness of
either  $\eps(M_1,\sfq)$ or  $d(M_1,\sfq)$ implies the finiteness of $\deq{M_1}$ \cite{BuPe07}.
Therefore, among the three measures of inaccuracy introduced above, the condition of
finite error bars gives the most general  criterion for selecting ``good" approximations of $\sfq$
and $\sfp$.

The following uncertainty relation for error bar widths is proven in \cite{BuPe06}.
\begin{theorem}\label{thm:err-bar-ur}
Let $M$ be an observable on $\brr$. The marginals $M_1,M_2$ obey the following
trade-off relation (for $0< \veps_1,\veps_2<\frac12$):
\begin{equation}\label{eqn:err-bar-ur}
\deq{M_1}\,\dep{M_2}\ge 2\pi\left( 1-\veps_1-\veps_2 \right)^2 \,\hbar.
\end{equation}
\end{theorem}

\subsubsection{Unsharpness}

There are various measures of the intrinsic unsharpness of an observable $E$ on $\br$.
Here we briefly review a measure based on the {\em noise operator} of $E$,
given by the positive operator $N(E):=E[2]-E[1]^2$. Note that this quantity appeared in the
definition of the standard
error, Eqs.~(\ref{eqn:stand-err02}), (\ref{eqn:glob-st-err}). The {\em (intrinisic) noise} is defined as
\begin{equation}\label{eqn:noise}
\mathcal{N}(E):=\sup_{T\in\sh}\tr{T\,N(E)}=\sup_{T\in\sh}\tr{T\,(E[2]-E[1]^2)}.
\end{equation}

In the case where $E[1]$ is a selfadjoint (rather than only symmetric) operator, it is known that
$N(E)=\nul$ if and only if $E$ is a sharp observable. The following trade-off relation for the noise
in approximate joint measurements of position and momentum is proven in \cite{BuPe07}.
\begin{theorem}\label{thm:noise-ur}
Let $M$ be an approximate joint observable for $\sfq,\sfp$ in the sense of finite error bars. Then
the noise of $M_1$ and the noise of $M_2$ obey the following inequality:
\begin{equation}
\mathcal{N}(M_1)\,\mathcal{N}(M_2)\ge\tfrac 12\hbar.
\end{equation}
\end{theorem}

An alternative measure of the intrinsic unsharpness of an observable on $\br$ is given by
the {\em resolution width}, introduced in \cite{CaHeTo06}; this quantity is similar in spirit to the
error bar width, and it is again found to yield a universal trade-off relation in joint measurements
\cite{BuPe07}.

\subsection{Inaccuracy-disturbance trade-off}

We have seen that a momentum measurement following a sharp position measurement defines
an observable that carries no information about the momentum distributions of the states prior to
the position measurement. A sharp measurement of position thus destroys completely the
momentum information contained in the initial state. The question arises whether  the disturbance
of momentum can be diminished if the position is measured approximately rather than sharply.

\begin{figure}
\begin{picture}(100,120)(-50,100)
\includegraphics[width=0.7\textwidth]{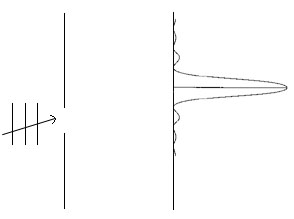} 
\end{picture}
\vspace{3cm}
\caption{Slit experiment as an approximate (sequential) joint measurement of position and
momentum.}\label{fig:slit}
\end{figure}

This possibility was already envisaged by Heisenberg in his discussion of thought experiments
illustrating the uncertainty relations \cite{Heisenberg27,PPQT}. For example, in the case of a
particle passing through a slit he noted that due to the diffraction at the slit,
an initially sharp momentum distribution is distorted into a broader distribution whose width
$\Delta p$ is of the order $\hbar/\delta x$, where  $\delta x$ is the width of the slit. The width
$\Delta p$ is a measure of the change, or disturbance, of the momentum distribution, and $\delta x$
can be interpreted as the inaccuracy of the position determination effected by the slit.
Further, one may also consider the recording of the location at which the particle hits
the screen as a geometric determination of the (direction) of its momentum, the inaccuracy
$\delta p$ of which is given by the width $\Delta p$ of the distribution obtained after many
repetitions of the experiment. In this way the passage through the slit
followed by the recording at the screen constitutes an approximate joint measurement of
the position and momentum of the particle at the moment of its passage through the slit;
see Figure \ref{fig:slit}.

Generalizing this idea of making an approximate joint measurement by way of a sequence of approximate measurements, we consider the schemes  of Figures \ref{fig:sharp-seq}
and \ref{fig:unsharp-seq}). Here $M_1$ is either the sharp position $\sfq$  or an unsharp
position observable $\sfq_e$ measured first, followed by a sharp momentum observable,
whose measurement
is to be followed by a sharp momentum measurement. The observable $M_2$ effectively
measured by this momentum measurement is defined via $\prob_{T}^{M_2}:=\prob_{T'}^\sfp$
for all initial states $T$, where $T'$ is the state after the position measurement. Thus $M_2$ is
the ``distorted" momentum observable. Collecting the probabilities for finding an outcome in a
set  $X$  for the first measurement and an outcome in $Y$ for the second measurement defines
a probability measure for each state $T$ via $X\times Y\mapsto \prob_T(X\times Y)$. Hence there is a unique
joint observable $M$ for $M_1$ and $M_2$ determined by the given measurement scheme \cite{BuHeLa06}. 

In the first case, since the marginal $M_1=\sfq$ is sharp, $M_2$ commutes with $\sfq$ and is
therefore {\em not} a good approximation of the momentum observable $\sfp$. However, in the
second case, $M_1=\sfq_e$, it is known \cite{Davies70} that the second marginal observable
$M_2$ is a smeared momentum observable, $M_2=\sfp_f$, if the first, unsharp position
measurement is such that the induced instrument is the von Neumann instrument
(\ref{eqn:vN-instrum}). The inaccuracy
distributions are then related as follows (cf. Eq.~(\ref{eqn:smeared-pos})):
\begin{equation}\label{eqn:pos-mom-inacc-distrib}
e(q)=\lambda|\phi(-\lambda q)|^2,\quad f(p)=\tfrac 1\lambda|\widetilde\phi(-\tfrac 1\lambda p)|^2.
\end{equation}
Here $\widetilde\phi$ is the Fourier transform of $\phi$, from which it follows that the standard deviations of the distributions $e,f$ obey the uncertainty relation:
\begin{equation}\label{eqn:seq-joint-mmt-ur}
\Delta(e)\Delta(f)\ge\tfrac 12 \hbar.
\end{equation}
Note that $\Delta(e)$, $\Delta(f)$ are measures of how well the sharp observables $\sfq,\sfp$ are
approximated by $M_1,M_2$, respectively. Thus they are measures of measurement inaccuracy,
and at the same time $\Delta(f)$ quantifies the disturbance of the momentum distribution due to
the position measurement.
\begin{figure}
{(1)} $M_1=\sfq$:
\[
\xymatrix{
T\ar[r]\ar[d]&\boxed{\sfq}\ar[r]\ar@{~>}[d]&T'\ar[r]&\boxed{\sfp}\ar@{~>}[d]\\
\prob_T^{\sfq},\prob_T^{\sfp}&\prob_T^{\sfq}=\prob_T^{M_1} &&
\prob_{T'}^{\sfp}=\prob_T^{f(\sfq)}=\prob_T^{M_2}
}
\]
\caption{Sharp position measurement followed by a sharp momentum measurement. The
two marginals $M_1=\sfq$ and $M_2$ commute and have a unique joint observable
$M$.}\label{fig:sharp-seq}
(2) $\xymatrix{M_1\ar@{~>}[r]&\sfq}$:
\[
\xymatrix{
T\ar[r]\ar[d]&\boxed{M_1}\ar[r]\ar@{~>}[d]&T'\ar[r]&\boxed{\sfp}\ar@{~>}[d]\\
\prob_T^{\sfq},\prob_T^{\sfp}&\prob_T^{M_1} &&\prob_{T'}^{\sfp}=\prob_T^{f(\sfq)}=\rho^{M_2}
}
\]
\caption{Approximate joint measurement of position and momentum defined by an unsharp position measurement followed by a sharp momentum measurement.
The marginals are $M_1=\sfq_e$ and $M_2=\sfp_f$ where $e,f$ are probability distributions
which are related as described in the main text.}\label{fig:unsharp-seq}
\end{figure}

These considerations show that an operational definition disturbance
of the momentum distribution due to a position measurement is
obtained by considering the sequential joint measurement  composed
of first measuring position and then momentum. The inaccuracy of the
second measurement, that is, any measure of the separation between
$\sfp$ and $M_2$, is also a measure of the momentum disturbance.
Consequently, all the joint measurement inaccuracy relations
discussed above apply to sequential joint measurements of position
and momentum, and in this case they  constitute rigorous versions of
the long-sought-after inaccuracy-vs-disturbance trade-off relations.

\section{Conclusion}\label{sec:conclusion}

Using the apparatus of modern quantum measurement theory, I have
reviewed rigorous formulations of some well-known quantum
limitations of measurements: the inevitability of disturbance and
(transient) entanglement; the impossibility of repeatable
measurements for continuous quantities, the restrictions on
measurements arising from the presence of an additive conserved
quantity, and the necessarily approximate and unsharp nature of
joint measurements of noncommuting quantities.

In each case, a strict no-go theorem is complemented with a
positive result describing conditions for an approximate realization
of the impossible goal: repeatability can be approximated
arbitrarily well for continuous sharp observables, also in the
presence of a conservation law. It was found that ideal measurements
of sharp observables are necessarily repeatable, but in the case of unsharp
observables, approximate ideality can be achieved without forcing 
approximate repeatability. Thus, unsharp measurements may be less invasive
than sharp measurements.

The impossibility of joint sharp measurements of complementary pairs
of observables can be modulated into the possibility of {\em approximate}
joint measurements of such observables, {\em provided} the inaccuracies
are allowed to obey a universal Heisenberg uncertainty relation. Likewise,
the complete destruction of momentum information by a sharp position 
measurement can be avoided if an {\em unsharp} position measurement is
performed. The trade-off between the
information gain in the approximate measurement of one observable and the
disturbance of (the distribution of) its complementary partner
observable was found to be an instance of the joint-measurement uncertainty
relation.

These results, some of which were made precise  in very recent investigations,
open up a range of interesting new questions and tasks. In particular, it will be important
to find operational measures of inaccuracy that are applicable to all types of observables,
whether bounded or unbounded, discrete or continuous. This would probably
enable a formulation of a universal form of joint measurement uncertainty relation
for arbitrary pairs of (noncommuting) observables, thus generalizing the relations
presented here for the special case of complementary pairs of continuous observables
such as position and momentum.

\vspace{6pt}

\noindent{\em Acknowledgement.} This work was carried out during my visiting appointment
at the Perimeter Institute (2005-2007). Hospitality and support by PI are gratefully acknowledged.




\end{document}